# Red Emission from Strain-Relaxed Bulk InGaN Active Region

Zuojian Pan, Zhizhong Chen*, Haodong Zhang, Chuhan Deng, Ling Hu, Fei Huang, Qi Wang, Guoyi Zhang, Xiaohang Li, and Bo Shen

[1] *State Key Laboratory for Artificial Microstructure and Mesoscopic Physics, School of Physics, Peking University, Beijing 100871, China*
*E-mail: zzchen@pku.edu.cn*
[2] *Advanced Semiconductor Laboratory, Electrical and Computer Engineering Program, CEMSE Division, King Abdullah University of Science and Technology (KAUST), Thuwal 23955-6900, Kingdom of Saudi Arabia*
[3] *Dongguan Institute of Optoelectronics, Peking University, Dongguan, Guangdong 523808, China*
[4] *Yangtze Delta Institute of Optoelectronics, Peking University, Nantong, Jiangsu 226000, China*

**Abstract**

High-In-content InGaN quantum wells (QWs) in red light-emitting diodes (LEDs) are typically grown at low temperatures to ensure effective In incorporation. In this study, red LEDs based on bulk InGaN active region were demonstrated. The growth temperature of bulk InGaN was ~800°C, which is over 100°C higher than the typical growth temperature of red QWs. By introducing high-density trench structures in the underlying green multi-quantum wells (MQWs), the compressive strain in bulk InGaN was relaxed by ~96%. With strain relaxation, phase separation occurred in the bulk InGaN, forming low-In-content (blue) and high-In-content (red) phases. The red phase acted as carrier localization centers, enabling red light emission under electrical injection. The red LEDs based on bulk InGaN exhibited a peak wavelength of 645 nm at 20 mA, with on-wafer peak external quantum efficiency of 0.32%. This study presents a new epitaxial strategy for red InGaN LEDs.

Phase separation is an intrinsic property of InGaN ternary alloys[1, 2]. It arises from the significant difference in covalent radii between InN and GaN, resulting in formation of a miscibility gap[1]. For InGaN grown on GaN under compressive strain, the miscibility gap shifts toward higher In compositions[3]. In fully strained states, InGaN films with In compositions up to 50% can be grown without phase separation[4, 5]. Strained high-In-content InGaN quantum wells (QWs) have been widely employed in red light-emitting diodes (LEDs)[6]. However, while strain suppresses phase separation, it simultaneously induces strong polarization fields in red QWs, leading to the quantum-confined Stark effect (QCSE)[7]. Additionally, the compressive strain hinders In incorporation in red QWs due to the high strain energy required for In-N bonds[8, 9].

Beyond strained InGaN QWs, strain-relaxed InGaN provides an alternative approach for red LEDs. Stringfellow *et al.* demonstrated that for strain-relaxed state, the solubility of InN in GaN is less than 6% at a typical growth temperature of 800 °C[1]. The separated phases typically consist of high-In and low-In regions[10-12]. An interesting observation is that the high-In phase exhibits higher radiative recombination efficiency compared to the low-In phase in the bulk InGaN[10, 13]. Moreover, the high-In phase has a narrower bandgap than the surrounding low-In phase, suggesting its potential as a carrier localization center. Thus, it is promising to use high-In phase in stress-relaxed InGaN as the active region to realize red InGaN LEDs.

In this study, red LEDs based on bulk InGaN active region were successfully demonstrated. By modulating the trench density in the underlying green MQWs, strained and relaxed bulk InGaN were obtained. Strained bulk InGaN showed almost no phase separation due to the stress-induced shift of the miscibility gap. In contrast, strain-relaxed bulk InGaN exhibited significant phase separation, forming low-In (blue) and high-In (red) phases. The red phase served as carrier localization centers, emitting red light under electrical injection. The red LED based on bulk InGaN achieved a peak wavelength of 645 nm at 20 mA, with an on-wafer peak external quantum efficiency (EQE) of 0.32%.

The epitaxial layers were grown on c-plane sapphire substrates using an Aixtron Crius I 31×2 inch close-coupled showerhead metal-organic vapor phase epitaxy (MOVPE) system. Figure 1(a) shows the LED structure based on bulk InGaN active region. The LED structure consists of a 1 μm unintentionally doped (UID) GaN layer, a 2 μm Si-doped GaN layer, six pairs of green MQWs, a 260 nm bulk InGaN active region, and a 160 nm p-GaN layer. The UID GaN, n-GaN, and p-GaN layers were grown in $H_2$ ambient using TMGa, $SiH_4$, $Cp_2Mg$, and $NH_3$, while the green MQWs and bulk InGaN were grown in $N_2$ ambient using TEGa, TMIn, and $NH_3$. The green MQWs consist of 2.5-nm InGaN QWs, 3-nm GaN cap layers, and 7-nm GaN quantum barrier (QB) layers, with the QWs and cap layers grown at the same temperature. As shown in Figure 1(b) and Table 1, two different growth temperatures were employed for green MQWs: sample A was grown at 850°C/980°C for the QW/QB layers, while sample B was grown at 760°C/830°C. Since the growth temperature affects In incorporation, the TMIn/TEGa ratio was adjusted to maintain a consistent emission wavelength: 140 sccm/700 sccm for sample A and 140 sccm/220 sccm for sample B. Apart from the green MQW growth conditions, samples A and B shared identical parameters for other layers. The bulk InGaN layers were grown at 940°C. The flow rate ratio of TMIn to TEGa was ~22:1, with a molar V/III ratio of about 3928. Meanwhile, Si gradient doping is applied during the bulk InGaN growth to reduce the turn-on voltage. The p-GaN layer comprised a low-temperature p-GaN layer grown at 920°C, a p-GaN and a $p^+$-GaN layer grown at 1050°C, and a p-InGaN contact layer grown at 810°C.

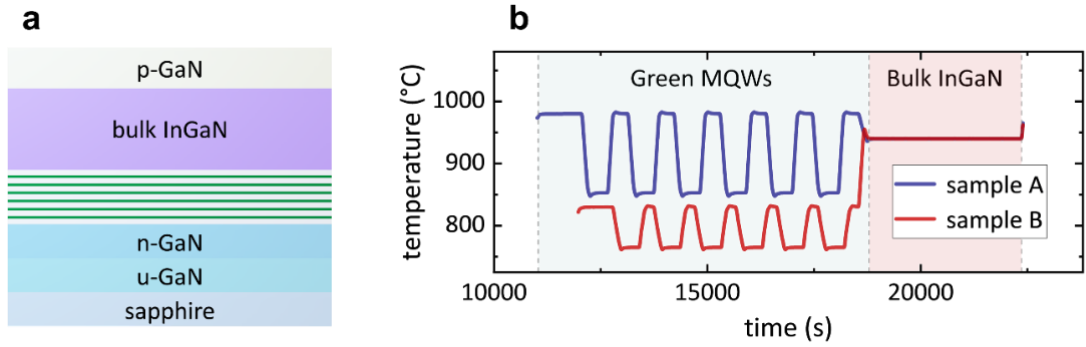

Figure 1. (a) Schematic diagram of the LED structure based on bulk InGaN active region. (b) Heater temperature profiles during the growth stages from green MQWs to bulk InGaN for samples A and B.

Table 1. Growth conditions and material properties of samples A and B

|  | Green MQWs | | | Bulk InGaN | |
| --- | --- | --- | --- | --- | --- |
|  | QW heater temp (wafer temp) | QB heater temp (wafer temp) | trench density | heater temp (wafer temp) | strain relaxation |
| Sample A | 850°C (~725°C) | 980°C (~843°C) | ~0 | 940°C (~802°C) | ~3 % |
| Sample B | 760°C (~657°C) | 830°C (~713°C) | ~5.6 ×10$^9$ cm$^{-2}$ | 940°C (~800°C) | ~96 % |

    Growth interruptions were performed after the green MQWs growth and bulk InGaN layers growth, respectively. The surface morphologies of samples A and B at different growth stages were measured by scanning electron microscope (SEM), as shown in Figure 2. In Figure 2(a), the green MQWs of sample A show a smooth surface with a low density of V-pits (~1.2×10$^8$ cm$^{-2}$). These V-pits mainly originate from dislocations in the underlying GaN. High growth temperature for green MQWs causes almost no trenches on the surface. In contrast, low-temperature grown green MQWs for sample B exhibit high-density trench structures (~5.6×10$^9$ cm$^{-2}$), as shown in Figure 2(b). These trench structures originate from high-density basal stacking faults (BSFs) in the green MQWs[14]. Stacking mismatch boundaries (SMBs) form around the BSFs. Meanwhile, V-pit loops generate along the SMBs, thus trench structures form finally [14]. In sample B, both the green QWs and QBs are grown at low temperatures. The reduced QW growth temperature promotes the formation of In-rich clusters, while the low QB growth temperature restricts the migration of Ga adatoms[15, 16]. The combination between In-rich clusters and the limited Ga adatom mobility resulted in the formation of high-density trench structures[15, 16].

    Figures 2(c) and 2(d) present the SEM images of the bulk InGaN surfaces for samples A and B. Notably, the growth conditions for the bulk InGaN in both samples were identical, with the only difference being the underlying green MQWs. In Figure 2(c), the bulk InGaN surface of sample A is relatively smooth, with a V-pit density of ~1.5×10$^8$ cm$^{-2}$, which is similar to that of green MQWs

in Figure 2(a). In contrast, the bulk InGaN surface of sample B exhibits significant roughness, with a high density of deep trench structures ~4.9×10⁹ cm⁻², as shown in Figure 2(d). The trench density is close to that of the green MQWs, but the trench depth is significantly increased. The deeper trenches in the bulk InGaN can be attributed to the expansion of V-pits, as the SMB from green MQWs extend upward. These high-density, deep trench structures in the bulk InGaN are likely to facilitate stress relaxation. From these observations, the bulk InGaN grown on high-temperature green MQWs maintains a relatively smooth surface, whereas that grown on low-temperature green MQWs exhibits a rough surface with a high density of deep trenches.

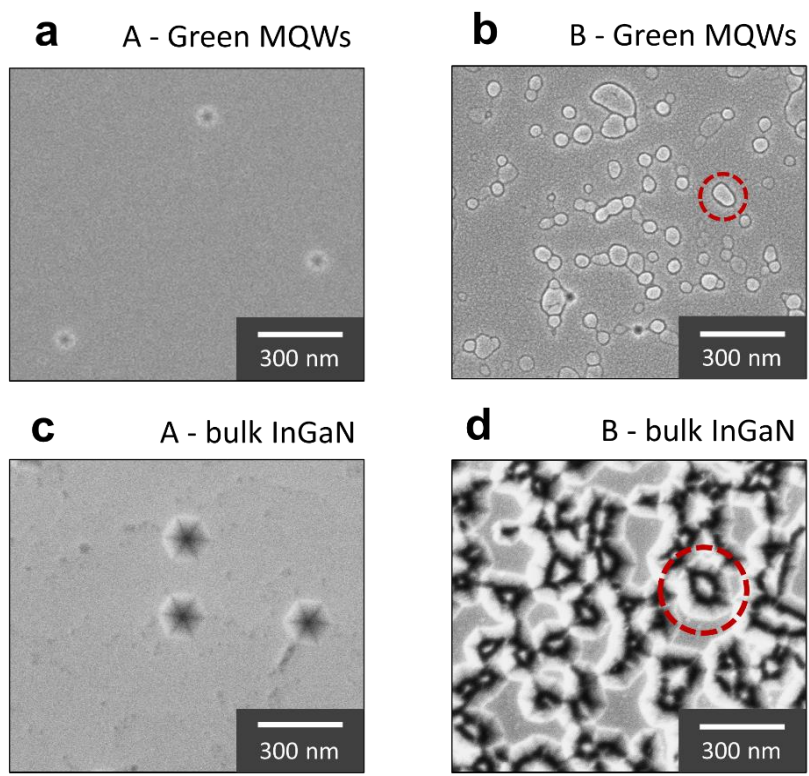

Figure 2. SEM images of (a) sample A and (b) sample B for green MQWs interruption, and (c) sample A and (d) sample B for bulk InGaN interruption.

X-ray diffraction reciprocal space mapping (XRD RSM) was measured to analyze the In content and stress states in the bulk InGaN layers. The In content was determined by combining the Poisson's ratios of GaN and InN with lattice spacing measurements along the (004) and (105) planes obtained from XRD RSM[17]. Figure 3 presents the XRD RSMs along the (105) reflection for both sample A and sample B. In Figure 3(a), the bulk InGaN layer in sample A exhibits an In content of 8.5%, with ~3% compressive strain relaxation. This suggests that V-pits in the bulk InGaN of sample A have a limited effect on compressive strain relaxation. In contrast, as shown in Figure 3(b), sample B demonstrates a higher In content of 11.9% in its bulk InGaN layer, with ~96% compressive stress relaxation. It indicates that the high density of deep trenches effectively relaxes the compressive stress and promotes the In incorporation in the bulk InGaN. Moreover, compared to sample A, the diffraction peak of bulk InGaN in sample B exhibits a broader distribution and weaker intensity, indicating more severe phase separation.

Strain directly influences the phase separation of InGaN[3, 4]. Figures 4(a-b) show the T-x phase diagrams of InGaN under strained and relaxed conditions[3]. In MQW-based LEDs, InGaN QWs are typically fully strained on the underlying GaN layer. For red LEDs, InGaN QWs are usually grown at temperatures between 600-700 °C with an In composition of 30-40%, as shown in Figure 4(a). The strong compressive stress in InGaN suppresses phase separation, ensuring high-quality epitaxy even with such high In compositions. For the strained bulk InGaN in sample A, the InGaN (802°C, In~8.5%) is located within the phase-stable region, thus avoiding phase separation, as marked in Figure 4(a). In contrast, for the relaxed bulk InGaN in sample B, the InGaN (grown at 800°C with ~11.9% In) lies near the binodal curve within the metastable phase region, as indicated in Figure 4(b). Thus, the bulk InGaN in sample B is highly susceptible to phase separation, leading to the formation of In-rich and In-poor phases.

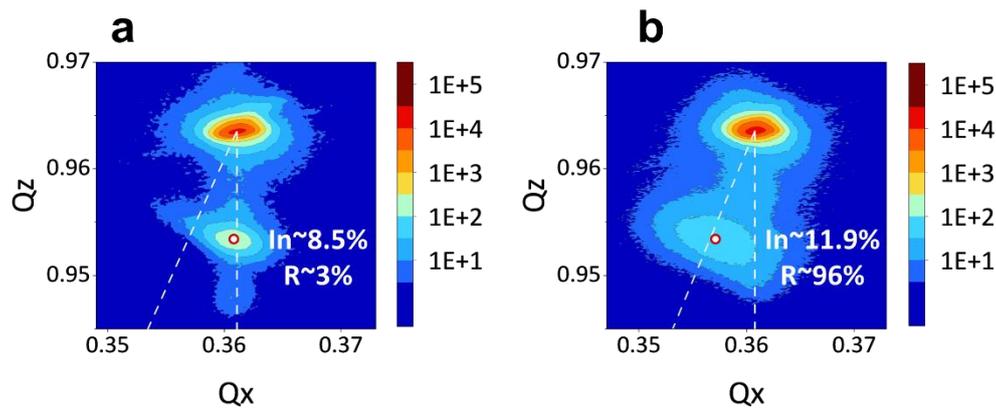

Figure 3. XRD RSM along the (105) reflection for (a) sample A and (b) sample B. The vertical and inclined dashed lines represent the fully strained and fully relaxed states, respectively.

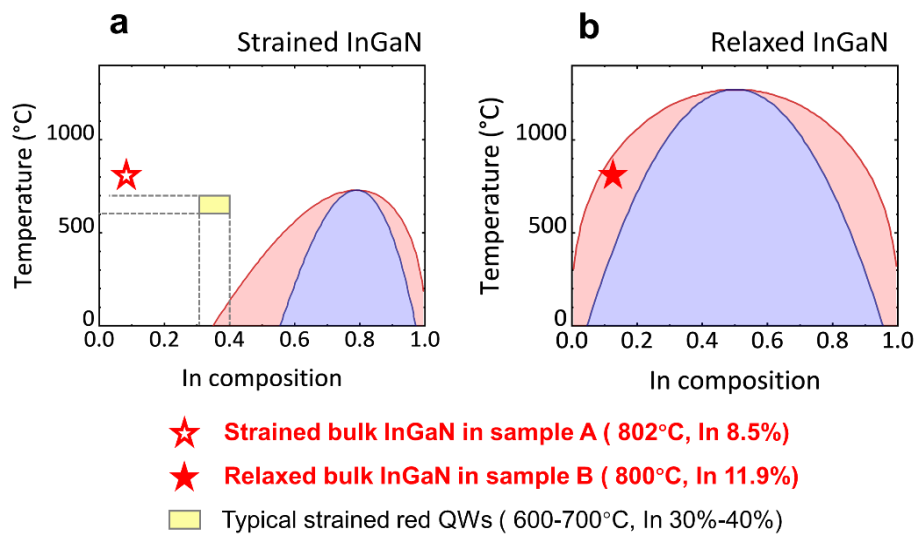

Figure 4. T-x phase diagrams of $In_xGa_{1-x}N$ compounds: (a) strained layers and (b) relaxed layers[3]. The positions of bulk InGaN in samples A and B, as well as typical strained red QWs, are indicated in the phase diagrams.

Cathodoluminescence (CL) was used to characterize the luminescence properties of bulk InGaN. The same acceleration voltage of 10 kV was applied to samples A and B. Figure 5(a) shows the CL spectrum of sample A, where there are two emission peaks at 392 and 506 nm. The two peaks originate from strained bulk InGaN and high-temperature-grown green MQWs, respectively. Interestingly, an unexpected red peak appears in the CL spectrum of sample B, as shown in Figure 5(b). The CL spectrum of sample B exhibits three emission peaks at 443, 501, and 635 nm. The 501 nm peak is attributed to low-temperature-grown green MQWs. The 443 nm blue peak and the 635 nm red peak are likely caused by phase separation in strain-relaxed bulk InGaN. Notably, bulk InGaN is grown at 800°C, which is over 100°C higher than the typical growth temperature of red QWs. At such a high temperature, In incorporation becomes highly challenging due to the high equilibrium vapor pressure of In-N[18]. However, in a strain-relaxed state, InGaN undergoes phase separation, forming high-In and low-In phases[10-12]. Thus, the high-In red phase is likely formed through phase separation in bulk InGaN rather than direct In incorporation. Tsai et al. reported that in thick InGaN films, a strain-relaxed high-In phase and a fully strained low-In phase can be observed[10]. Due to carrier localization, the high-In phase exhibits higher radiative recombination efficiency and longer carrier lifetime than the low-In phase[10]. In Figure 5(b), the red peak exhibits significantly higher CL intensity than the blue peak in sample B, likely due to carrier localization. The red phase, with a narrower bandgap than the blue phase, acts as carrier localization center, demonstrating its potential as the active region for red LEDs.

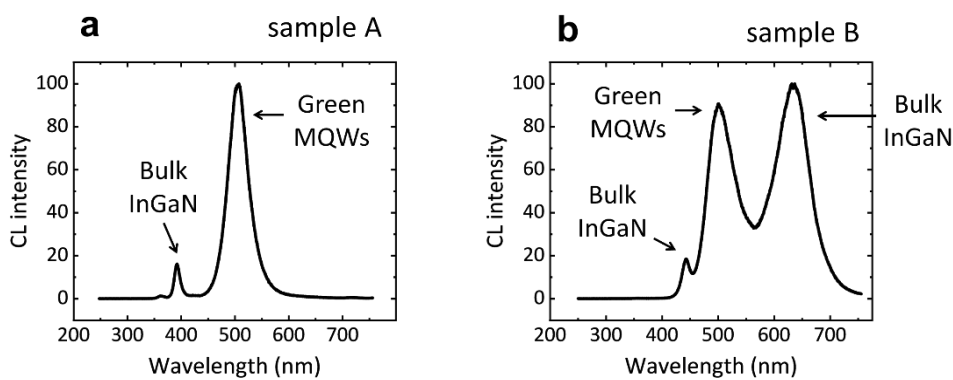

Figure 5. CL spectra of bulk InGaN in (a) sample A and (b) sample B at 10 kV.

Electroluminescence (EL) measurements were performed on the LED wafer of sample B, using In balls as electrical contacts. An integrating sphere collected light from the LED wafer's backside to measure the emission spectrum and light output power. As shown in Figure 6(a), sample B exhibits a single red emission peak at 645 nm with a full width at half maximum (FWHM) of 68 nm under a 20 mA injection current. It means that the red phase in bulk InGaN achieves effective red light emission under electrical injection. The I-V curve in Figure 6(b) indicates a threshold voltage of ~3.2 V for sample B. Generally, a thick undoped bulk InGaN layer results in a high operating voltage. To mitigate this issue, we introduced a graded Si doping profile in the bulk InGaN layer, effectively reducing the operating voltage. Figure 6(c) presents the on-wafer EQE and wall plug efficiency (WPE) curves of sample B. At a 1 mA injection current, the EQE and WPE reach 0.32% and 0.29%, respectively. As the current increases, efficiency gradually decreases due to Auger recombination and carrier overflow. For MQW-based LEDs, the QW width of 2–3 nm provides excellent carrier confinement, ensuring high carrier injection efficiency. However, bulk InGaN lacks effective carrier confinement, and the low carrier injection efficiency may result in a reduced EQE. Further structural optimization is necessary to enhance the efficiency. Overall, the high-In red phase separated from bulk InGaN was used as the active region for the first time, achieving red light emission under electrical injection.

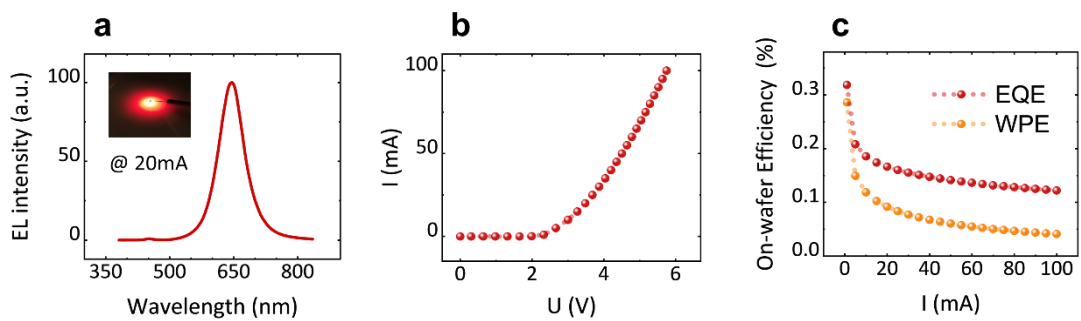

Figure 6. EL performance of sample B using In balls as contacts: (a) EL spectrum at a current of 20 mA, with the inset showing the corresponding EL image. (b) I-V curve. (c) On-wafer EQE and WPE curves measured from 1 mA to 100 mA.

In summary, red LEDs based on bulk InGaN active region were successfully demonstrated. Bulk InGaN grown on high-temperature green MQWs remains nearly fully strained without significant phase separation. In contrast, bulk InGaN grown on low-temperature green MQWs with high-density trench structures exhibits ~96% strain relaxation, approaching a fully relaxed state. CL results reveal that strain-relaxed bulk InGaN promotes phase separation, forming blue and red InGaN phases. The red phase acts as carrier localization centers, enabling red light emission under electrical injection. The red LED with bulk InGaN active region exhibits a peak wavelength of 645 nm at 20 mA, with an on-wafer peak EQE of ~ 0.32%. This study presents a novel epitaxial approach for red InGaN LEDs.

**Acknowledgements**
This work was supported by the National Key Research and Development Program of China (2021YFB3600100, 2022YFB3608501, 2023YFB4604401), National Natural Science Foundation of China (62174004, 61927806, 81871427), Basic and Applied Basic Research Foundation of Guangdong Province (2020B1515120020).

**Data Availability Statement**
The data that support the findings of this study are available from the corresponding author upon reasonable request.

**Conflict ofInterest**
The authors declare no conflict of interest.